\documentclass[twocolumn,showpacs,preprintnumbers,amsmath,amssymb]{revtex4}
\usepackage{tabularx,graphicx}

\usepackage{color}
\usepackage{hyperref}
\hypersetup{
    colorlinks=true,
    linkcolor=blue,
    filecolor=blue,      
    urlcolor=blue,
}

\begin{document}

\newcommand{\beq}{\begin{equation}}
\newcommand{\eeq}{\end{equation}}
\newcommand{\beqn}{\begin{eqnarray}}
\newcommand{\eeqn}{\end{eqnarray}}
\newcommand{\bmath}{\begin{subequations}}
\newcommand{\emath}{\end{subequations}}
\newcommand{\bra}[1]{\langle #1|}
\newcommand{\ket}[1]{|#1\rangle}

\title{Moment of inertia of superconductors
}
\author{J. E. Hirsch }
\address{Department of Physics, University of California, San Diego\\
La Jolla, CA 92093-0319}
 
\begin{abstract} 

We find that the bulk moment of inertia per unit volume of a metal becoming superconducting increases by the amount
$m_e/(\pi r_c)$, with $m_e$ the bare electron mass and $r_c=e^2/m_e c^2$ the classical electron radius.
This is because superfluid electrons acquire an intrinsic moment of inertia 
$m_e (2\lambda_L)^2$, with $\lambda_L$ the London penetration depth.
As a consequence, we predict that when a rotating long cylinder  becomes superconducting its angular velocity does not change,
contrary to the prediction of conventional BCS-London theory that  it will rotate faster.  We explain the dynamics
of magnetic field generation when a rotating normal metal becomes superconducting. \end{abstract}
\maketitle

  \begin{figure}
\resizebox{8.5cm}{!}{\includegraphics[width=7cm]{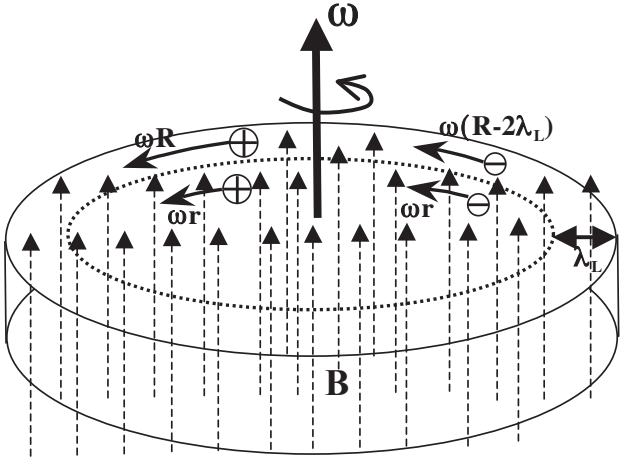}}
  \caption{Cylinder rotating with angular velocity  $\omega$. In the interior, at radial distance $r$ from the rotation axis,
  ions and electrons move with the same speed $\omega r$. Near the  surface, ions move with speed $\omega R$  and
  electrons move with slower speed $\omega(R-2\lambda_L)$. A magnetic field $B$ exists throughout the interior
  (dashed arrows),
  generated by the surface current resulting from   electrons  lagging behind the  ions.}
\end{figure}

\section{introduction}
A superconducting body  rotating with angular velocity $\vec{\omega}$ develops a uniform magnetic field throughout its interior, given by
\beq
\vec{B}=-\frac{2m_ e c}{e}\vec{\omega} \equiv B (\omega)\hat{\omega}
\eeq
with $e$ ($<0$) the electron charge, and $m_e$ the bare electron mass. 
Numerically, $B=1.137\times 10^{-7} \omega$, with $B$ in Gauss and $\omega$ in rad/s. The magnetic field is parallel to the angular velocity. 
The phenomenon was predicted in 1933, just before the Meissner effect was discovered, 
by Becker, Heller and Sauter \cite{becker} for a perfect conductor set into rotation. Subsequently London predicted 
that the same final state should result when a rotating normal metal is cooled into the superconducting state \cite{londonbook}.
The resulting magnetic moment will depend on the shape of
the body and is called the ``London moment''. In this paper we propose that this effect reveals fundamental physics of 
superconductors not predicted by conventional BCS theory \cite{schrieffer,tinkham}. For a preliminary 
treatment where we argued that BCS  theory is inconsistent with the London moment
we refer the reader to our earlier work \cite{lm1}. Other non-conventional explanations of the London moment have also been 
proposed \cite{tajmar,tajmar2}.

Eq. (1) has been verified experimentally for a variety of superconductors \cite{hild,brick,
lm2,lm3,lm4,lm5,lm6,lm7}. The same result is obtained whether the sample is first cooled and then set into
rotation, or cooled while rotating, as predicted by London \cite{londonbook}. 
Rotation speeds used in experiment are typically of order 20 to 100 revolutions/second. 
Equation (1) follows from London's
equation \cite{tinkham}
\beq
\vec{\nabla}\times\vec{v}_s=-\frac{e}{m_ec}\vec{B} .
\eeq
with $\vec{v}_s$ the superfluid velocity. In the interior of the body, the superfluid rotates together with the body, hence the superfluid velocity
  at distance $r$ from the rotation  axis is given by
\beq
\vec{v}_s=\vec{\omega}\times\vec{r}
\eeq
and using $\vec{\nabla}\times (\vec{\omega}\times\vec{r})=2\vec{\omega}$ in Eq. (2), Eq. (1) results.

Consider the simplest geometry, a long cylinder of radius R, where no demagnetization effects exist. A section of such a cylinder is
shown in Figure 1. For a supercurrent  flowing within the 
London penetration depth ($\lambda_L$) of the surface, the superfluid velocity that excludes an applied magnetic
field $\vec{B}$ is  
\beq
\vec{v}_s=-\frac{e\lambda_L}{m_e c}\vec{B} .
\eeq
Therefore, to generate the interior London field Eq. (1) in a rotating superconductor
in the absence of applied external field, the superfluid
velocity in a rim of thickness $\lambda_L$ at the surface has to be
\beq
\vec{v}_s=\vec{\omega}\times\vec{r} (1-\frac{2\lambda_L}{R})
\eeq
so the superfluid  velocity in the rim lags the rotation of the body by the small amount 
\beq
\Delta v_s=-2\lambda_L\omega .
\eeq
The full behavior of
superfluid velocity versus radius is given by Laue \cite{laue} for a cylindrical geometry and by
Becker et al \cite{becker} and London  \cite{londonbook} for a spherical geometry.
The London penetration depth is given by \cite{tinkham}
\beq
\frac{1}{\lambda_L^2}=\frac{4\pi n_se^2}{m_e c^2}
\eeq
with $n_s$ the superfluid density.

For a perfect conductor that is set into rotation, Eq. (1) follows from Maxwell's equations, as discussed by Becker et al \cite{becker}, assuming the superfluid electrons are completely detached from the lattice. As the body is set into
rotation the moving ions generate an electric  current and hence a time-dependent magnetic field that generates a Faraday electric
field that pushes the superfluid electrons to follow the motion of the ions, albeit with a small lag near the surface
that gives rise to the surface current that generates the magnetic field Eq. (1). The derivation is reviewed in Appendix A.

In this paper we explain the behavior of rotating superconductors using the theory of hole superconductivity \cite{holesc},
and we argue that   the conventional theory of superconductivity is inconsistent with the physics of rotating superconductors.

\section{the conventional view}
We consider a normal metallic cylinder rotating around its axis with angular velocity $\vec{\omega}$ that is cooled into
the superconducting state. We assume the cylinder is floating in a gravitation-free environment and there is no friction.
Therefore, total angular momentum is conserved. 

In the normal state, electrons and ions rotate together at the same speed. When the cylinder becomes superconducting, electrons near the surface
have to spontaneously slow down to the velocity Eq. (5) upon entering the superconducting state. 
The conventional theory of superconductivity
does not explain the dynamics of this  process, which London himself considered  to be   ``quite absurd'' from the perfect conductor viewpoint \cite{absurd} and did not provide an explanation for.
Furthermore,  the body has to spontaneously speed up to compensate for the electronic angular
momentum change. The conventional theory of superconductivity does not explain what is the physical process that causes the
body to increase its rotation velocity.

The total mechanical angular momentum of electrons plus ions is
\beq
\vec{L}=\vec{L}_e+\vec{L}_{i} .
\eeq
The electronic angular momentum in the rotating normal state  is
\bmath
\beq
L_e=I_e\omega
\eeq
\beq
I_e=\frac{\pi}{2}n_sm_e h R^4
\eeq
\emath
where $h$ is the height of the cylinder. $I_e$ is the moment of inertia of the superfluid electrons. As the system goes superconducting, electrons in the rim slow down according to Eq. (5), and the electronic angular momentum decreases by
\beq
\Delta L_e=(2\pi R \lambda_L h n_s)\times (m_e \Delta v_s R)
\eeq
with $\Delta v_s$ given by Eq. (6). In Eq. (10), the first factor is the number of superfluid electrons in the surface rim, and the second factor is the change in
angular momentum of one electron. Using Eq. (7), Eq. (10) can be rewritten as
\beq
\Delta L_e=-(\frac{m_e c }{e})^2 R^2h\omega .
\eeq

The kinetic energy of the superfluid electrons in the normal state due to the body rotation is
\beq
K_e=\frac{1}{2} I _e \omega^2 =\frac{L_e^2}{2I_e}
\eeq
and when the system goes superconducting and the electrons in the rim slow down  it decreases by the amount
\beq
\Delta K_e=\frac{L_e\Delta L_e}{I_e}=\Delta L_e \omega
\eeq
or, using Eqs. (11) and (1)
\beq
\Delta K_e=-\frac{B^2}{4\pi}(\pi R^2 h) .
\eeq
For the ions, the    kinetic energy of rotation is
\beq
K_i=\frac{L_i^2}{2I_i}
\eeq
with $I_i$ the moment of  inertia for the ions. From Eq. (8) and conservation of total angular momentum we deduce
\beq
\Delta L_i=-\Delta L_e,
\eeq
therefore, angular momentum has to be transferred from the electrons to the body. This gives rise to a change in the
kinetic energy of the ions,  
\beq
\Delta K_i= \frac{L_i}{I_i}\Delta L_i =\Delta L_i\omega=-\Delta K_e 
\eeq
hence
\beq
\Delta K_e+\Delta K_i=0.
\eeq
Therefore,  conservation of angular momentum requires that the change in ionic and electronic kinetic energies of rotation
exactly compensate each other. The body's rotation speed slightly increases to compensate for the slowing down of the
rim electrons, and its kinetic energy of rotation slightly increases. The conventional theory of superconductivity does not explain how the kinetic energy saved by the lagging electrons is transferred to the body to make it speed up.
The frequency of rotation when the rotating normal metal becomes superconducting will be larger than $\omega$ by the amount
\beq
\Delta \omega=\frac{\Delta L_i}{I_i}=\frac{m_e}{m_pA}\frac{n_s}{n}\frac{8\lambda_L^2}{R^2}\omega
\eeq
where $m_p$ is the nucleon mass, A the atomic weight and $n$ the ionic number density.

For a type I superconductor it is expected \cite{babaev} that when the body is rotating at frequency $\omega$ the
transition to the superconducting state will occur when the system is cooled below $T_c$ to the temperature $T$ for which 
\beq 
H_c(T)=-\frac{2m_e c}{e}\omega=B .
\eeq
with $H_c(T)$ the thermodynamic critical field.
Now  the magnetic field energy associated with the  magnetic field Eq. (1)   for a
cylinder with zero demagnetizing factor  is given by
\beq
E_B=\frac{B^2}{8\pi}(\pi R^2 h)=\frac{1}{2}|\Delta K_e|.
\eeq
The condensation energy for a superconducting cylinder at rest is given by
\beq
E_{cond}=\frac{H_c(T)^2}{8\pi}n_s(\pi R^2 h) .
\eeq
In the Meissner effect, the condensation energy provides the energy required to expel the magnetic field.
Here, the condensation energy Eq. (22) provides the energy to generate the magnetic field Eq. (21).
Therefore, conservation of energy requires that the condensation energy Eq. (22) is the same for
rotating and non-rotating superconductors.

However, we argue that for a superconductor rotating at high speeds it is inconsistent to assume that the binding
energy of Cooper pairs, that determines the condensation energy, would be independent of rotation speed.
According to the conventional theory this should be true for arbitrarily high frequencies, even for
frequencies giving rise to a magnetic field $B$ larger than the thermodynamic critical field
at zero temperature.   If we consider two  electrons in a Cooper pair separated by a typical distance
$\xi \sim \lambda_L$, the energy associated with rotation at frequency $\omega$ is
\beq
\epsilon_{rot}=m_e\lambda_L^2\omega^2 \sim B^2/(8\pi n_s)
\eeq
i.e. the magnetic field energy per electron. We argue that it is inconceivable that the binding energy of Cooper pairs would
not be lowered  by the rotation energy $\epsilon_{rot}$ for any magnitude of $B$. 

However, if the condensation energy is decreased at finite rotation frequency,  there is not enough energy to account for the creation of the magnetic field Eq. (1) when the system
becomes superconducting, because the energy saved in the slowing down of the rim electrons was entirely used up in 
speeding up the body to satisfy angular momentum conservation. Therefore, we argue that the conventional theory cannot account for the physics
of rotating superconductors without violating  either conservation of energy or conservation
of angular momentum.

The implausibility of the conventional picture  also  follows from the following argument. From the Meissner effect we learned  that the final 
state of a superconductor in a magnetic field $B$ is unique, independent of history. 
Whether we apply a magnetic field $B$ to a normal metal and then cool it to the superconducting state, or we apply the
same magnetic field $B$ to a metal already superconducting, the final state of the system is exactly the same.
Similarly here,  the rotation frequency $\omega$ plays the role that  $B$ plays in the Meissner effect. Whether
we apply the rotation $\omega$ to the normal metal and then cool it to the superconducting state, or instead apply
the same rotation $\omega$ to a metal already superconducting, the final state of the system should be exactly the same, with
the same $\omega$ \cite{absurd}. This implies that {\it the rotation speed of the body  should not change when  the rotating metal enters the superconducting state}, contrary to the prediction of the conventional theory that the rotation frequency
should change by the amount given by Eq.  (19). 

Finally, as reviewed in Appenix A, for the case of a superconductor at rest that is  set into rotation, the derivation of Eq. (1)
requires that there is no direct interaction between the  superfluid
electrons and the lattice. If that is the case, how can there be a transfer of angular momentum from
electrons to the body as a rotating normal metal becomes superconducting?

\section{alternative view}

Instead, we propose that when the rotating cylinder becomes   superconducting its angular velocity {\it does not  change} and
hence that the ionic angular momentum doesn't change.  This then implies that the total 
electronic angular momentum does not  change either.

The angular momentum of electrons in the rim decreases by the amount Eq. (10). Hence  the angular momentum of
electrons in the bulk has to increase by the same amount:
\bmath
\beq
\Delta L_e + \Delta L_e^{bulk}=0
\eeq
\beq
\Delta L_e^{bulk}=m_e(2\lambda_L)^2n_s(\pi R^2h) \omega
\eeq
\emath
Since the number of electrons in the bulk is
 \beq
N_e= n_s\pi R^2h 
\eeq
this implies that each electron in the bulk acquires an additional  `intrinsic' angular momentum
\beq
\ell=m_e (2\lambda_L)^2 \omega.
\eeq

Eq. (26) says that when the rotating metal becomes superconducting there is an additional contribution to its electronic
angular momentum that comes from the electron mass being spread out in a ring of radius $2\lambda_L$.  Equivalently, that the electron's orbit expands from a microscopic radius to radius $2\lambda_L$.
 Such physics  was predicted by the theory
of hole superconductivity  \cite{sm,bohr} to explain the dynamics of the Meissner effect: as an electron
expands its orbit from point-like to radius $2\lambda_L$ in the presence of a magnetic field, azimuthal current of the magnitude required to expel
the magnetic field gets generated by the Lorentz force. Expansion of the electronic wavefunction to radius $2\lambda_L$ is an indispensable element in the dynamical explanation of the Meissner effect within
this theory \cite{meissnerexp}. In Eq. (26) we find a direct confirmation of this essential part of the theory.  It implies that 
when electrons enter the superfluid state they acquire an {\it intrinsic moment of inertia}
\beq
i_{el}=m_e (2\lambda_L)^2.
\eeq
so the total electronic moment of inertia increases by
\beq
\Delta I_e=m_e (2\lambda_L)^2n_s(\pi R^2 h)
\eeq
and the electronic angular momentum increase in the bulk Eq. (24b) is
\beq
\Delta L_e^{bulk}=\Delta I_e \omega
\eeq
Using Eq. (7), we can write the increase in moment of inertia as 
\beq
\Delta I_e=m_e\frac{V}{\pi r_c}
\eeq
where $V=\pi R^2h$  is the volume of the body and $r_c=e^2/m_ec^2$ is the classical electron radius. Because
Eq. (30) no longer depends on the geometry of  the body  we believe it is very likely that it is a general result  
for a body of arbitrary shape and for any rotation axis.

Note the interesting fact that Eq. (30) is also independent  of the superfluid density and the London penetration depth.
Thus it expresses a $qualitative$ difference between tne normal and superconducting states of matter.
Presumably the volume factor in Eq. (30) corresponds to the volume of the sample that is  in the superconducting state.

How does the energetics work in this scenario? We have from Eqs. (13) and (14)  that the decrease in electronic kinetic energy
because of the rim slowing down is
\beq
\Delta K_e=\Delta L_e \omega = -\Delta I_e \omega^2=-\frac{B^2}{4\pi}(\pi R^2 h)
\eeq
The ions do not
acquire extra kinetic energy since the frequency of rotation $\omega$ doesn't change. The bulk electronic kinetic energy is given by Eq. (12), so when the moment of inertia increases it increases by
\beq
\Delta K_e^{bulk}=\frac{1}{2} \Delta I_e \omega^2 =\frac{B^2}{8\pi}   (\pi R^2 h)
\eeq
which is half of the decrease in rim kinetic energy Eq. (31). The other half goes into paying the cost in magnetic energy Eq. (1).
This then implies that the condensation energy of the rotating superconductor at temperature $T$  is, instead of Eq. (22)
\beq
E_{cond}(T,\omega)= (\frac{H_c^2(T)}{8\pi}- \frac{B^2(\omega)}{8\pi})(\pi R^2 h)
 \eeq
 with $B(\omega)$ given by Eq.(1). The
  transition occurs at the frequency or temperature where the condensation energy vanishes.

\section {kinetics  of the transition}

Let us consider the process by which the magnetic field attains the value Eq. (1). Consider a point at distance
$r$ from the origin. From Faraday's law
\beq
\oint \vec{E}\cdot \vec{dl}=-\frac{1}{c}\frac{\partial \phi}{\partial t}
\eeq
and assuming cylindrical symmetry we have
\beq
E_F(r,t)=-\frac{1}{2\pi r} \frac{1}{c}\frac{\partial \phi (r,t)}{\partial t}
\eeq
where $E_F(r,t)$ is the induced Faraday  electric field and $\phi(r,t)$ the magnetic flux through $r'<r$, at time $t$. 
Integrating over time,
\beq
\int_0^\infty dt E_F(r,t)=\frac{B}{2cr}=\frac{m_e}{e}\omega r .
\eeq
 If the location $r$ is superconducting, a superfluid electron at $r$ obeys the equation of motion
 \beq
 \frac{dv_s}{dt}=\frac{e}{m_e} E(r,t)
 \eeq
 where $v_s$ and $E$ are in the azimuthal direction.  Assuming the point $r$ was superconducting
 during the entire process we have upon integrating Eq. (37) and using Eq. (36)
 \beq
 \Delta v_s=\int_0^\infty  \frac{dv_s}{dt} = \omega r .
 \eeq
 Therefore, this equation describes the process shown in Fig. 2 (a),  where the body is initially at rest in the superconducting
 state and the electron initial speed is $v_s(R,t=0)=0$, and attains final speed $\omega r$ when
 the body acquires angular velocity $\omega$. Eq. (38) does not apply when $r$ is within
 $\lambda_L$ of the surface because the magnetic field does not acquire its full bulk value
 Eq. (1) in that region.

  \begin{figure}
\resizebox{9.0cm}{!}{\includegraphics[width=7cm]{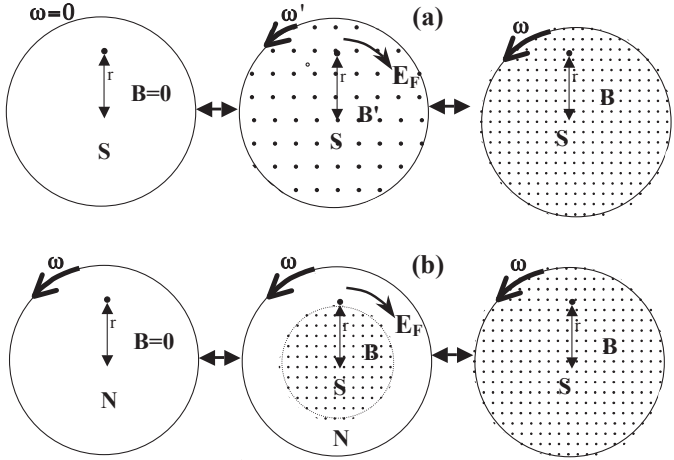}}
  \caption{
  Growth of the magnetic field in rotating superconductors (from left to right). (a) A superconductor at rest starts rotating until it reaches
  frequency $\omega$. (b) A normal metal (N) enters the superconducting state (S) upon cooling, while rotating at frequency $\omega$ .
  Dots indicate magnetic field pointing out of the paper, their density indicates the strength of the field. At a given point $r$ in the interior, the change
  in magnetic flux through the region $r'<r$   between the initial and final states is 
  given by $\Delta \phi= \pi r^2 B$ and determines the time integral of the Faraday electric field at point $r$,   Eq. (36).}
\end{figure} 
 
 This reasoning also shows that if we are considering the process where the rotating normal metal
 is cooled into the superconducting state, an electron at radius $r$ has to be in the normal
 state during the entire time where the magnetic flux $\phi(r,t)$ changes. This is because initially electrons at radius $r$ in the normal state rotate
 together with the body with azimuthal speed $\omega r$. If at any time while $\phi$ is changing Eq. (37) was valid,
 the electron at $r$ would attain a final velocity different from $\omega r$, however we know that 
 in the final state electrons in the interior rotate together with the body with azimuthal speed $\omega r$.
 We conclude that when a rotating metal is cooled into the superconducting state the superconducting
 region necessarily expands from the inside out, as shown schematically in Fig. 2 (b). 
 The Faraday field acts on electrons at point $r$ in the normal state while the magnetic field is
 growing in the region $r'<r$, during this time it generates Joule heat but does not change the electron's azimuthal 
 speed.
 
   \begin{figure}
\resizebox{8.5cm}{!}{\includegraphics[width=7cm]{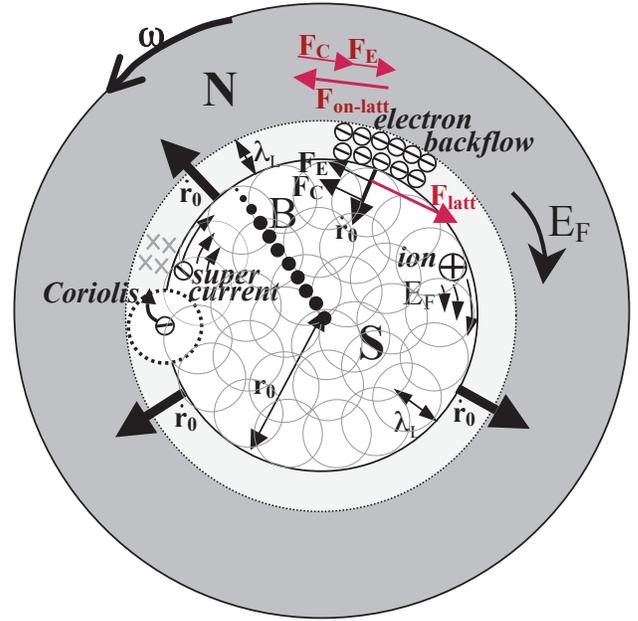}}
  \caption{Rotating normal metal becoming superconducting.
  The phase boundary at radius $r_0$ expands outward. Black dots indicate magnetic field pointing out of the paper, grey crosses
  indicated demagnetizing field pointing into the paper (see Sect. VI). Electrons at the phase boundary expand 
  their orbits to radius $2\lambda_L$, and a backflow of normal electrons takes place to compensate for the
  radial charge imbalance.}
\end{figure} 

  \begin{figure}
\resizebox{8.5cm}{!}{\includegraphics[width=7cm]{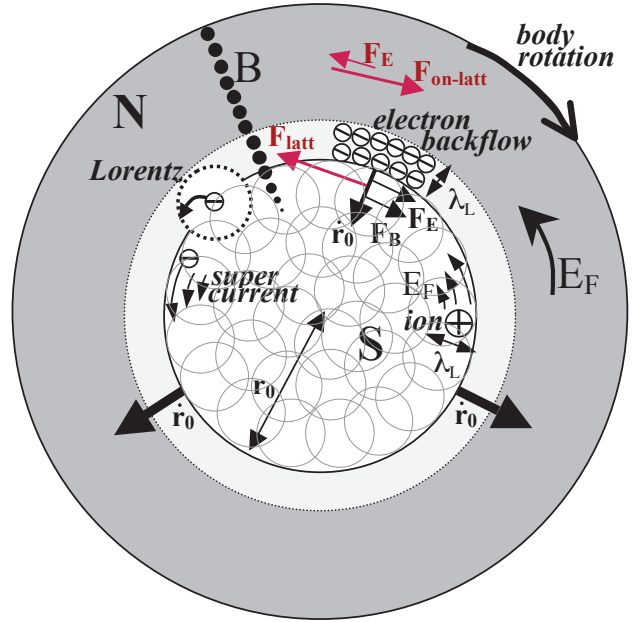}}
  \caption{Meissner effect.  See text and ref. \cite{momentum} for details.}
\end{figure} 

 \section {dynamics of the transition}
 Next we analyze the dynamics of the transition when  the rotating normal metal becomes
 superconducting, along the same lines that we analyzed the 
 dynamics of the Meissner effect in Refs. \cite{momentum,spinning}. There are many similarities but some 
 important subtle differences. Figures 3 and 4 show schematically both processes,
 where $r_0$ denotes the radius of the expanding phase boundary. The magnetic field
 pointing out of the paper is denoted by black circles, with their diameter illustrating the intensity.

 The driving force for the generation of the surface current in both cases is expansion of the electronic
 orbits from a microscopic radius to radius $2\lambda_L$ when the electrons enter the
 superconducting state. In the Meissner case, this expansion in the
 presence of magnetic field $B$ produces the Meissner counterclockwise current  through the action of the
 magnetic Lorentz force. Similarly, in the rotating case the Coriolis force acts
 \beq
 \vec{F}_C= 2m_e \vec{v}\times \vec{\omega}=-\frac{e}{c}\vec{v}\times\vec{B}(\omega) 
 \eeq
 imparting clockwise speed (relative to the body) to the outgoing electron with radial velocity $\vec{v}$.

 In the Meissner effect, the body acquires momentum in the opposite direction through the backflow
 process illustrated in Fig. 4: backflowing electrons {\it with negative effective mass}  are subject to a force from the ions in the 
 counterclockwise direction ($F_{latt}$), causing a clockwise reaction force to be exerted on the body $F_{on-latt}$ that makes
 the body turn. That force is partially compensated by the counterclockwise force exerted on the positive ions
 by the Faraday field. The quantitative analysis is given in ref. \cite{momentum}.
 
 The momentum transferred by the backflowing electrons to the body in the Meissner effect can be thought of as momentum
 that was stored in the electromagnetic field when electrons expanded their orbits and acquired azimuthal momentum
 through the Lorentz force \cite{momentum}. Therein lies the essential difference with the case of the rotating superconductor:
 there, when the orbit expands and acquires azimuthal momentum through the Coriolis force (Fig. 3), no momentum
 is stored in the electromagnetic field because there is no magnetic field in the lightly shaded region in Fig. 3
 for a long cylinder. Instead,
 the same Coriolis force that deflects the electron expanding its orbit in the clockwise direction transfers momentum to the
 ions in counterclockwise direction. That momentum is cancelled by the clockwise momentum transferred by the
 Faraday electric field $E_F$ to the ions in that region. 
 
 The balance of forces for the backflowing electrons for the rotating superconductor  is shown in Fig. 3. Coriolis and electric forces act counterclockwise,
 exactly balanced by the clockwise force exerted on the electrons by the lattice of ions ($F_{latt}$) so that backflowing
 electrons flow radially in, just like for the Meissner effect. However unlike for the Meissner case, here the backflowing electrons
 do not transfer net momentum to the body because the reaction to the force exerted by the ions on the electrons,
 $F_{on-latt}$, is cancelled by the sum of electric force $F_E$ and the reaction to the Coriolis force on the electrons
 $F_C$ (Fig. 3). 
 
 In summary, for the reasons explained above, unlike for the Meissner effect there is no net transfer of
 angular momentum to the body for the rotating superconductor as the superconducting region expands.
 The fact that backflowing electrons have negative effective mass of course still plays an essential role:
 if they had positive effective mass, backflowing electrons would be deflected counterclockwise and
 transfer their momentum to the body by collisions. This would of course not alter the 
 angular momentum balance compared to the scenario described above, but would cause 
 dissipation and entropy production rendering the transition irreversible in contradiction with 
 theory and experiment \cite{momentum,entropy}.

  \section{non-zero demagnetizing factor}
  For samples other than long cylinders a demagnetizing magnetic field will exist, and the above considered simplest situation 
  needs to be modified. Consider for example the case of a rotating sphere \cite{becker,londonbook}.
  The lagging velocity of electrons in the rim is not constant as given by Eq. (6) but rather 
  depends on the location of electrons relative to the equator \cite{londonbook}:
  \beq
\Delta v_s=3\lambda_L (sin \theta) \omega
\eeq
where $\theta$ is the angle between the position vector and the axis of rotation. So at the equator,
the lagging speed of electrons is larger than for the cylinder by a factor $3/2$. How can this be understood
within our scenario, and what are its consequences?
  
For the case of the Meissner effect (Fig. 4), it is immediately clear why the electrons in the Meissner current at the
equator acquire the higher speed: the magnetic field that imparts the azimuthal speed through the Lorentz force
acting on the expanding orbits is larger than the applied field precisely by the factor $3/2$, due to 
demagnetization. Recall that the critical magnetic field for a sphere is $(2/3)H_c$ rather than $H_c$ \cite{tinkham}.

Similarly we can understand the larger lagging speed for the spherical rotating superconductor. In Fig. 4, in the lightly shaded
region where the electron orbit  expands  and is deflected clockwise by the Coriolis force, there is now a magnetic
field pointing $into$ the paper because of demagnetization (indicated by crosses in Fig. 3). The magnetic  Lorentz force on the expanding orbit 
provides additional azimuthal momentum in the clockwise direction in addition to the one imparted by the Coriolis force. It also stores some momentum
in the electromagnetic field. The angular momentum density in the electromagnetic field is
\beq
\vec{L}_{em}=\frac{1}{4\pi c} \vec{r}\times(\vec{E}\times \vec{B}).
\eeq
The electric field $\vec{E}$ created by the outflow of negative charge resulting from orbit enlargement points radially outward,
the demagnetizing field $\vec{B}$ points into the paper, hence $\vec{L}_{em}$ points parallel to $\vec{\omega}$ (out of the paper)
and will increase the angular velocity when transferred to the body by the backflowing electrons. 
Thus, unlike the case of the long cylinder where there is no demagnetizing field, here (and for any sample with non-zero demagnetizing factor)
there will be a small change (increase) in the velocity of rotation of the body when the rotating
normal metal becomes superconducting.

For example, for the case of the sphere of radius $R$ the angular momentum of the rim current is
\beq
\Delta L_e=2(\frac{m_e c }{e})^2 R^3 \omega
\eeq
and the increase  in the moment of inertia that we predict is, from Eq. (30)
\beq
\Delta I_e=\frac{4}{3} (\frac{m_e c }{e})^2 R^3
\eeq
so it only accunts for $2/3$ of the angular momentum change Eq. (42). Thus the angular velocity
will increase slightly by $\Delta \omega$ given by
\beq
\Delta \omega =\frac{2(\frac{m_e c }{e})^2 R^3}{3I}\omega
\eeq
with $I$ the moment of inertia of the body, while in the conventional theory the increase in rotation frequency is given by
\beq
\Delta \omega =\frac{2(\frac{m_e c }{e})^2 R^3}{I}\omega
\eeq
i.e. a factor of $3$ larger. We explained above how the body acquires the additional rotation speed within our theory.
  
    \section{experimental consequences}
    Our theory predicts that upon cooling a rotating long cylinder into the superconducting state, its rotation frequency should not change.
    Instead, the conventional theory predicts the body's rotation frequency should increase by the amount given by Eq. (19). For Al, with $A=27$,
    density $\rho=2.7 g/cm^3$, $\lambda_L(T=0)=500 \AA$, this is
    \beq
    \Delta \omega=3.05 \times 10^{-5} (\frac{\lambda_L}{R})^2 \omega
    \eeq
    Unfortunately, even for very thin cylinders with radius approaching $\lambda_L$, this small change would be extremely
    difficult to detect both by direct measurement of the frequency or by measuring the resulting very small
    change in the magnetic field, $\Delta B=1.137\times 10^{-7}\Delta \omega$. .
    
    What about directly  measuring the change in moment of inertia between normal and superconducting states?
        Using a sensitive torsional oscillator, experiments attempting to measure  `non-classical rotational inertia' in solid
    $^4 He$ were performed to study possible `supersolid' behavior \cite{chan}.  
    For rotating superconductors we predict that the total moment of inertia should not change between normal and superconducting states, because
    the increase in bulk moment of inertia is compensated  by lowering of the moment of inertia of the rim. Instead,
    the conventional theory predicts an overall decrease in the moment of inertia upon entering the superconducting state.
    However, again the predicted change is extremely small:
    \beq
    \frac{\Delta I}{I}=-\frac{\Delta \omega}{\omega}
    \eeq
    so that its detection is probably beyond the capabilities of  even the most sensitive torsional oscillators.

We also  predict that at the transition point the condensation energy goes to zero (Eq. 33), while the
    conventional theory predicts it is finite (Eq. 22). Correspondingly, we predict no latent heat and the conventional
    theory predicts finite latent heat. Within the two-fluid model the latent heat per unit volume at the transition
    temperature $T$ according to the
    conventional theory is
    \beq
    L(T)=\frac{H_0}{2\pi} (\frac{T}{T_c})^2 H_c(T) \sim \frac{H_0 B(\omega)}{2\pi}
    \eeq
    with $H_0$ the zero temperature critical field. For $H_0=500G$,
    \beq
    L(T)=0.91\times 10^{-5} \omega (rad/s) erg/cm^3
    \eeq
which is extremely small even for very high rotation frequencies, hence very difficult to measure.

\section{discussion}

In this paper we have proposed that when a rotating normal metal becomes superconducting the rotation speed of 
the body does not change, contrary to what the conventional theory predicts, when the electrons near the surface slow
down relative to the body motion and create the interior magnetic field Eq. (1). This unexpected effect occurs because electrons increase their
contribution to the bulk moment of inertia of the body when they enter the superconducting state according to the
theory of hole superconductivity, thus increasing the bulk electronic  angular momentum and thereby 
compensating for the decrease in the surface electrons angular momentum. 
Each superfluid electron acquires an intrinsic angular momentum Eq. (26) that adds to the total angular momentum of the body
without increasing its angular velocity.

Our scenario certainly resolves the question  of  how to explain a speed-up of the rotational velocity of a normal metallic cylinder
when it   becomes superconducting, for which the conventional theory has provided no mechanism: there is no speedup. 
We also explain the mechanism that causes   the rim electrons to slow down when the rotating normal metal becomes superconducting.
In a nutshell, the Coriolis force acting on expanding orbits. The conventional theory provides no explanation, it simply postulates that it happens.

In fact, we found that except in the case of zero demagnetizing factor our theory also predicts a speedup of rotation
when the normal metal becomes superconducing, albeit by a smaller amount than predicted by the conventional theory.
We   explained how the demagnetizing field leads to a larger decrease in the speed of
surface electrons and to transfer of angular momentum to the body, using the same concepts that we recently used
to explain the dynamics of the Meissner effect, in particular the fact that normal state charge carriers have to
be hole-like \cite{momentum}.

Within the theory of hole superconductivity, electrons expand their orbits from microscopic radius $k_F^{-1}$ 
($k_F$= Fermi momentum) to radius $2\lambda_L$ when they pair up and become superconducting \cite{sm}.
This explains why the diamagnetic susceptibility grows from the normal metal Landau susceptibility
to $-1/(4\pi)$ \cite{emf}, since the Larmor diamagnetic susceptibility is proportional to the square of the radius
of the orbit. In addition this provides a pictorial understanding of the development of macroscopic phase
coherence due to overlapping orbits \cite{emf}. The fact that the moment of inertia of the body
 increases by $m_e r_0^2$ per electron when the orbit of the
electron increases from essentially zero radius to radius $r_0=2\lambda_L$ follows from simple geometry illustrated in Fig. 5. Fig, 5
shows a point-like particle labeled 1  at radius $r$ from the center, and another point-like particle labeled 2 that
moves in an  orbit of radius $r_0$ centered at distance $r$ from the center. Simple geometry shows that the
moment of inertia of particle 1 relative to an axis going through the center of the large circle is 
$mr^2$ and that of particle 2 is $m(r^2+r_0^2)$ ($m$=mass of the particles) assuming particle 2 is uniformly distributed along its orbit of
radius $r_0$. 

  \begin{figure}
\resizebox{5.5cm}{!}{\includegraphics[width=7cm]{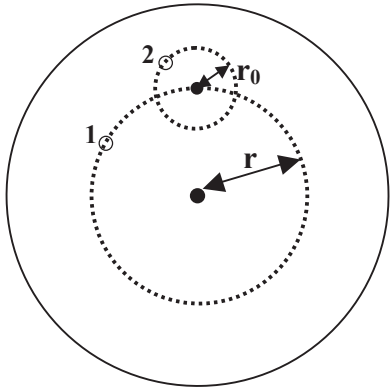}}
  \caption{Geometric illustration of increase in moment of inertia caused by orbit enlargement (see text).}
\end{figure}

The reason why the increase in the bulk moment of inertia of the body Eq. (30) is independent of the direction of the rotation axis is
easy to understand. When the body rotates, the expanded orbits will orient themselves so that they lie
on planes perpendicular to the rotation axis to minimize their energy, as classical particles would.

It was pointed out by Bethe long ago  \cite{bethe} that the moment of inertia of the electronic charge distribution in a solid 
determines its mean inner electric potential. Shortly thereafter, Rosenfeld  \cite{rosenfeld} pointed out that the
mean inner potential is proportional to the diamagnetic susceptibility.
In other work we have pointed out \cite{mip} that the same physics discussed here, increase in electronic moment of inertia
when a normal metal becomes superconducting, should lead to an increase in both its diamagnetic susceptibility
(as observed) and in its mean inner potential, which can be measured by electron holography \cite{holography} (not yet observed).

Within the theory of hole superconductivity, superconductors have a macroscopically inhomogeneous charge distribution in their
ground state, with more negative charge near the surface \cite{chargeexp}. One may wonder whether this will give an additional effect
at sufficiently low temperatures where the resulting electric field is not screened by normal quasiparticles (that would 
also cancel the mass imbalance). This was suggested in Ref. \cite{lm1}.  In fact, it will not.
The expelled electrons will rotate together with the body and the speed of body rotation will not change. 
The extra angular momentum of the expelled electrons will be exactly compensated by angular momentum
stored in the electromagnetic field and the angular momentum balance discussed in this paper will not change.

We have pointed out that there is an inconsistency in the conventional theory, that assumes that there is no direct interaction between electrons and
the body when a superconductor at rest is set into rotation (Appendix A), yet requires a transfer of momentum between electrons and the
body when a rotating normal metal becomes superconducting. How is that inconsistency resolved in our theory? In
our theory, there can be a net momentum transfer between electrons and the body when there is $radial$ charge flow. Such charge flow
occurs in the process of the normal metal becoming superconducting, but not when a superconductor is set into rotation.
We can explain both situations where there is no momentum transfer and where there is, provided in the latter case
there is also radial charge flow.
Within the conventional theory, there never is radial charge flow.

We point  out that the physics discussed here is closely related to physics  of superfluid 
$^4He$. $^4He$ has maximum density at the superfluid  transition temperature  \cite{onnesexpansion}. Below the superfluid transition the system $expands$ when cooled further.
As a consequence  its bulk moment of inertia will increase as it enters the superfluid state, as found here
for superconductors. We have argued
elsewhere that this commonality between superconductors and superfluid $^4He$ derives from the fact that
$both$ superconductivity and superfluidity are {\it kinetic energy driven}, originating in expansion of the
wavefunction driven by quantum pressure \cite{helium,helium2}. The physics of the Meissner effect discussed in
refs. \cite{meissnerexp,momentum}
involving flow of superfluid and counterflow of normal fluid is also closely related to physics 
found in $^4He$ that gives rise to the fountain effect \cite{londonbook2}. We conjecture that the concepts discussed here may be relevant to the understanding of
rotation experiments in superfluid $^4He$ \cite{osborne,andron,fairbank}.

In summary we find that superconducting matter has a new distinct property.   In addition to its electrical conductivity becoming infinite and its magnetic susceptibility 
becoming that of
a perfect diamagnet, a metal becoming superconducting will increase   its bulk  moment of inertia per unit volume
  by
the amount $(m_ec/e)^2/\pi=1.03\times 10^{-15}g/cm$.
This  extra  moment of inertia arises from the development of an intrinsic moment of inertia
for each superfluid electron. Superfluid electrons behave  as  an extended rim of mass $m_e$,  radius
$2\lambda_L$, intrinsic moment of inertia $m_e (2\lambda_L)^2$, and intrinsic orbital
angular momentum $\hbar/2$ \cite{bohr} rather than as  point  particles.
This leads to a dynamical understanding of the Meissner effect and the London moment,
and follows from the quantization of orbital angular momentum in the presence of 
Dirac's spin orbit interaction predicted by the theory \cite{bohr}.
Unfortunately as we saw in Sect. VII  it appears very difficult in practice to test the different predictions of our theory
vis-a-vis the conventional theory for rotating superconductors experimentally. 

\appendix

\section{Derivation of Eq. (1) from Maxwell's equations}
When a superconducting body with cylindrical symmetry  is set into rotation, the superfluid electrons obey the equation of motion
\beq
m_e\frac{dv_s}{dt}=eE+F_{latt}
\eeq
where the first term is the force on the electrons from the induced azimuthal electric field, the second term is a direct
force that may be exerted by the ions on the superfluid electrons, and $m_e$ is the $bare$ electron mass. The electric field $E$ is determined by
Faraday's law Eqs. (34) and (35), which yield at radial position r
\beq
E(r,t)=-\frac{1}{2c}r \frac{dB(r,t)}{dt} .
\eeq
In the interior, superfluid electrons rotate together with the body, i.e. with azimuthal velocity $v_s=\omega r$. 
Using this and combining Eqs. (A1) and (A2) then yields
\beq
m_e r \frac{d \omega}{dt}=--\frac{e}{2c}r \frac{dB(r,t)}{dt}+F_{latt}.
\eeq
Under the assumption that $F_{latt}=0$, Eq. (A3) integrates to
\beq
m_e \omega=-\frac{e}{2c}B
\eeq
i.e. Eq. (1), assuming the initial conditions are $\omega=B=0$. 

In other words, Eq. (1) is obtained for a superconductor set into rotation {\it under the assumption}
that there is no direct force $F_{latt}$ acting between the ions and electrons, i.e. that the electrons
are perfectly free from interactions with the ions. This is precisely what was assumed in the original work by Becker et al \cite{becker}:
that ``die mittlere Geschwindigkeit der Elektronen nur unter der Wirkung
eines elektrischen Feldes \"{a}ndern'' (``the mean velocity of the electrons only changes under the effect of an electric field'').


\begin{references} 


\bibitem{becker} R. Becker, G. Heller und F. Sauter,
 `\"Uber die Stromverteilung in einer supraleitenden Kugel', 
 \href{http://link.springer.com/article/10.1007/BF01330324}
 {Zeitschrift f\"ur Physik  {\bf 85}, 772 (1933).}
\bibitem{londonbook}  F. London, ``SuperfluidsÕ', Volume I, Wiley, New York, (1950).

\bibitem{schrieffer} J.R. Schrieffer, ``Theory of Superconductivity'', Westview Press, Boulder, 1999. 
\bibitem{tinkham}   M. Tinkham,  ``Introduction to Superconductivity'', 2nd ed, McGraw Hill, New York, 1996.
\bibitem{lm1} J. E. Hirsch, 
``The London moment: what a rotating superconductor reveals about superconductivity'',
\href{http://iopscience.iop.org/article/10.1088/0031-8949/89/01/015806/meta} {Physica Scripta {\bf 89}, 015806 (2013)}.

\bibitem{tajmar} M. Tajmar, ``Electrodynamics in superconductors explained by Proca equations'',
\href{https://www.sciencedirect.com/science/article/pii/S0375960107015551}
{Phys. Lett. A {\bf 372}, 3289 (2008).}

\bibitem{tajmar2} A. K. T. Assis and M. Tajmar, 
``Superconductivity with Weber's electrodynamics: The London moment and the Meissner effect'',
\href{http://aflb.ensmp.fr/AFLB-422/aflb422m864.htm}{Annales de la Fondation Louis de Broglie {\bf 42}, 307 (2017)}.  

\bibitem{hild} A.  F. Hildebrandt,
`Magnetic Field of a Rotating Superconductor',
\href{http://journals.aps.org/prl/abstract/10.1103/PhysRevLett.12.190}
{Phys. Rev. Lett. {\bf 12}, 190 (1964).}
\bibitem{brick} N. F. Brickman, ``Rotating Superconductors'', 
\href{https://journals.aps.org/pr/abstract/10.1103/PhysRev.184.460}{Phys. Rev. {\bf 184}. 460 (1969)}.
\bibitem{lm2} A.F. Hildebrand and M.M. Saffren, in Proceeding of the 9th International
Conference on Low Temperature Physics, edited by
J.G. Daunt et al. Plenum, New York, 1965, p. 459.
\bibitem{lm3} M. Bol and W.M. Fairbank, in Proceedings of the 9th International
Conference on Low Temperature Physics, edited by
J.G. Daunt et al. Plenum, New York, 1965, p. 471.
\bibitem{lm4} J. Tate, B. Cabrera, S. B. Felch and J.T. Anderson, 
\href{https://journals.aps.org/prl/abstract/10.1103/PhysRevLett.62.845}{Phys. Rev. Lett. {\bf 62}. 845 (1989)}.
\bibitem{lm5} D. Hipkins, W. Felson and Y.M. Xiao, 
\href{https://link.springer.com/article/10.1007/BF02570422}{Czech. J. Phys. {\bf 46}, Suppl. S5, 2871 (1996)}.
\bibitem{lm6} A.A. Verheijen et al., Physica B 165-166, 1181 (1990).
\bibitem{lm7} M. A. Sanzari,H. L. Cui, and F.  Karwacki, \href{https://aip.scitation.org/doi/10.1063/1.116622}{Appl. Phys. Lett. {\bf 68}, 3802 (1996)}.

  \bibitem{laue} M. von Laue, ``Theory of Superconductivity'', Academic Press, New York, 1952.

 \bibitem{holesc} References in   \href{http://sdphln.ucsd.edu/~jorge/hole.html}{http://physics.ucsd.edu/$\sim$jorge/hole.html}.
   
   \bibitem{absurd} London's remarks \cite{londonbook}: 
{\it There is an implication which might be worth mentioning since it would appear quite absurd from the viewpoint of the perfect conductor concept. The uniqueness properties of the present
 theory provide for only $one$ current distribution {\it independent of the way} in which the superconducting state is reached. Consequently we have to conclude that the {\it same state} as has been
 calculated above must also be obtained if the sphere is cooled from the normal into the superconducting state while it is rotating. The perfect conductor theory would, of course,
 furnish no reason for the electrons near the surface of the metal to lag suddenly behind when the rotating sphere goes into the superconducting state. 
 It would simply lead to a state of zero magnetic moment in which all charges move in phase - the same below as above the transition point.}
 
 \bibitem{babaev} E. Babaev and B. Svistunov, 
 ``Rotational response of superconductors: Magnetorotational isomorphism
and rotation-induced vortex lattice'',
 \href{https://journals.aps.org/prb/abstract/10.1103/PhysRevB.89.104501}{Phys. Rev. B {\bf 89}, 104501 (2014).}
   
       \bibitem{sm} J.E. Hirsch,  
    ``Spin Meissner effect in superconductors and the origin
of the Meissner effect'',
       \href{http://iopscience.iop.org/0295-5075/81/6/67003}{Europhys. Lett. {\bf 81}, 67003 (2008)}.
                \bibitem{bohr} J. E. Hirsch, ``The Bohr superconductor'',  \href{http://iopscience.iop.org/article/10.1209/0295-5075/113/37001}
    {Europhys. Lett. {\bf 113}, 37001 (2016)}.
    
     
 \bibitem{meissnerexp} J. E. Hirsch, ``On the dynamics of the Meissner effect'', 
 \href{http://iopscience.iop.org/article/10.1088/0031-8949/91/3/035801}
 {Physica Scripta {\bf 91}, -05801 (2016)}.
    
      \bibitem{momentum} J.E. Hirsch,  ``Momentum of superconducting electrons and the explanation of the Meissner effect'',
   \href{http://journals.aps.org/prb/abstract/10.1103/PhysRevB.95.014503}{Phys. Rev. B {\bf 95}, 014503 (2017)}.
   
    \bibitem{spinning} J.E. Hirsch,  ``Spinning Superconductors and Ferromagnets'',
   \href{http://przyrbwn.icm.edu.pl/APP/ABSTR/133/app133-3-5.html}{Acta Physica Polonica A {\bf 133}, 350 (2018)}.
   

     
 
     

    
      \bibitem{entropy} J.E. Hirsch, 
     ``Entropy generation and momentum transfer in the superconductor to normal phase transformation and the consistency of the conventional theory of superconductivity'', \href{https://www.worldscientific.com/doi/10.1142/s0217979218501588}.   
    
        \bibitem{chan} D. Y. Kim and M. H. W. Chan,
     \href{https://journals.aps.org/prl/abstract/10.1103/PhysRevLett.109.155301}{Phys. Rev. Lett. {\bf 109}, 155301 (2012)}.
     
     


\bibitem{emf} J.E. Hirsch, `Electromotive Forces and the Meissner Effect Puzzle',
\href{http://link.springer.com/article/10.1007%2Fs10948-009-0531-4}{J. Sup. Nov. Mag. {\bf 23}, 309 (2010)}.


\bibitem{bethe} H. Bethe, ``Theorie der Beugung von Elektronen an Kristallen'',
\href{https://onlinelibrary.wiley.com/doi/abs/10.1002/andp.19283921704}{Ann. Physik {\bf 392}, 85 (1928)}. 

\bibitem{rosenfeld} L. Rosenfeld, ``Brechungsindex der Elektronen und Diamagnetismus'',
\href{https://link.springer.com/article/10.1007%2FBF01506319?LI=true}    {Naturwissenschaften {\bf 17}, 49 (1929)}.

\bibitem{mip} J. E. Hirsch, ``Superconductivity, diamagnetism, and the mean inner potential of solids'',
\href{http://onlinelibrary.wiley.com/doi/10.1002/andp.201300147/abstract} {Annalen der Physik
{\bf 526}, 63 (2013)}.

\bibitem{holography} A. Tonomura, ``Electron Holography'', Springer, Berlin, 1999, and references therein.


 \bibitem{chargeexp} J.E. Hirsch,  `Charge expulsion and electric field in
superconductors',  \href{http://journals.aps.org/prb/abstract/10.1103/PhysRevB.68.184502}{Phys.Rev. B {\bf 68}, 184502 (2003)}.

\bibitem{onnesexpansion} 
H. K. Onnes, ``Further experiments with liquid helium. A'', in:
KNAW, Proceedings, 13 II, 1910-1911, Amsterdam, 1911, pp. 1093-1113;
H. K. Onnes and J. D. A. Boks, ``The variation of density of liquid helium below the boiling point'',
Reports and Communications of the Fourth International Congres of Refrigeration, June 1924,
Comm. 170b, p 18-23.


  \bibitem{helium} J.E. Hirsch,  ``Kinetic energy driven superconductivity and superfluidity'', 
  \href{https://www.worldscientific.com/doi/abs/10.1142/S0217984911027613}{Mod. Phys. Lett. B 25, 2219 (2011)}.
  \bibitem{helium2} J. E. Hirsch, ``Kinetic energy driven superfluidity and superconductivity and the origin
of the Meissner effect'', \href{http://www.sciencedirect.com/science/article/pii/S0921453413001172}{Physica C {\bf 493}, 18 (2013)}.

\bibitem{londonbook2}   F. London, ``Superfluids'', Vol. II, Dover, New York, 1964.

\bibitem{osborne} D. V. Osborne, \href{http://iopscience.iop.org/article/10.1088/0370-1298/63/8/315}
{Proc. Phys. Soc. {\bf A 63},  909 (1950)}.
\bibitem{andron}  E. L. Andronikashvili and Y. G. Mamaladze, 
\href{https://journals.aps.org/rmp/abstract/10.1103/RevModPhys.38.567}{Rev. Mod. Phys. {\bf 38}, 567 (1966)}.
\bibitem{fairbank} G. B. Hess and W. M. Fairbank,
\href{https://journals.aps.org/prl/abstract/10.1103/PhysRevLett.19.216}{Phys. Rev. Lett. {\bf 19}, 216 (1967)}.


 \end{references}
 \end{document}